\documentclass[12pt]{article}
\usepackage{epsfig}
\usepackage{rotating}
\newcommand{\be}{\begin{equation}}
\newcommand{\ee}{\end{equation}}
\newcommand{\ba}{\begin{eqnarray}}
\newcommand{\ea}{\end{eqnarray}}
\newcommand \w {{\omega}}
\newcommand \Cov {{\rm Cov}}

\newcommand \Dt {{\Delta t}}

\begin{document}

\begin{center}
\Large{\textbf{Endogenous versus Exogenous Origins of Crises}}\\~\\
\normalsize{D. Sornette} 
\end{center}
\begin{center}
\normalsize{
Institute of Geophysics and Planetary Physics\\ and
Department of Earth and Space Sciences\\
University of California, Los Angeles, California 90095, USA\\
Laboratoire de Physique de la Mati\`{e}re Condens\'{e}e CNRS UMR 6622\\
Universit\'{e} de Nice-Sophia Antipolis, 06108 Nice Cedex 2, France}
\normalsize{\today}\\~\\
\end{center}

\begin{abstract}
Are large biological extinctions such as the Cretaceous/Tertiary KT
boundary due to a meteorite, extreme volcanic activity or
self-organized critical extinction cascades? Are commercial successes
due to a progressive reputation cascade or the result of a well
orchestrated advertisement? 
Determining the chain of causality for extreme events in complex systems
requires disentangling interwoven exogenous and endogenous
contributions with either no clear or too many signatures. Here, I review
several efforts carried out with collaborators, which suggest a general strategy for understanding
the organization of several complex systems under the dual
effect of endogenous and exogenous fluctuations. The studied examples are:
Internet download shocks, book sale shocks, social shocks, financial volatility shocks,
and financial crashes. Simple models are offered to quantitatively relate the
endogenous organization to the exogenous response of the system. 
Suggestions for applications of these ideas to many other systems are offered.
\end{abstract}

\section{Introduction}

Extreme events are pervasive in all natural and social systems:
earthquakes, volcanic eruptions, hurricanes and tornadoes, landslides
and avalanches, lightning strikes, magnetic storms, catastrophic events
of environment degradation, failure of engineering structures, crashes
in the financial stock markets, social unrests leading to large-scale
strikes and upheavel and perhaps to revolutions, economic drawdowns on
national and global scales, regional and national power blackouts,
traffic gridlocks, diseases and epidemics and so on.

Can we forecast them, manage, mitigate or prevent them? The answer
to these questions requires us to investigate their origin(s). 

Self-organized criticality, and more generally, complex system theory
contend that out-of-equilibrium slowly driven systems with threshold
dynamics relax through a hierarchy of avalanches of all sizes.
Accordingly, extreme events are seen to be endogenous \cite{BakPak,Bak}, in contrast with
previous prevailing views. In addition, the preparation processes
before large avalanches are almost
undistinguishable from those before small avalanches, making the prediction
of the former basically impossible (see \cite{Srpredicat} for a discussion).
But, how can one assert with 100\% confidence
that a given extreme event is really due to an endogenous
self-organization of the system, rather than to the response to an
external shock? Most natural and social systems are indeed continuously
subjected to external stimulations, noises, shocks, sollications,
forcing, which can widely vary in amplitude. It is thus not clear a
priori if a given large event is due to a strong exogenous shock, to the
internal dynamics of the system organizing in response
to the continuous flow of small sollicitations, or maybe to a combination of both.
Adressing this question is fundamental for understanding the relative importance of
self-organization versus external forcing in complex systems and for the 
understanding and prediction of crises.

This leads to two questions:
\begin{enumerate}
\item Are there distinguishing properties that
characterize endogenous versus exogenous shocks?

\item  What are the relationships between endogenous and exogenous shocks?
\end{enumerate}
Actually, the second question has a long tradition in physics. It is 
at the basis of the interrogations that scientists perform on the enormously
varied systems they study. The idea is simple: subject the system to 
a perturbation, a ``kick'' of some sort, and measure its response as a function
of time, of the nature of the sollicitations and of the various environmental
factors that can be controlled. In physical systems at the thermodynamic
equilibrium, the answer is known under the name of the theorem of fluctuation-dissipation, 
sometimes also refered to as the theorem of fluctuation-susceptibility \cite{stratfluc}.
In a nutshell, this theorem relates quantitatively in a very precise way
the response of the system to an instantaneous kick (exogeneous) to the correlation function
of its spontaneous fluctuations (endogenous). An early example of this relationship is
found in Einstein's relation between the diffusion coefficient $D$ of 
a particle in a fluid subjected to the chaotic collisions of the fluid molecules
and the coefficient $\eta$ of viscosity of the fluid \cite{Einstein1,Einstein2}. 
The coefficient $\eta$ controls the drag,
i.e., response of the particle velocity when subjected to an exogenous force impulse.
The coefficient $D$ can be shown to be a direct measure of the (integral of the) 
correlation function of the spontaneous (endogenous) fluctuations of the particle velocity.

In out-of-equilibrium systems, the existence of a relationship between 
the response function to external kicks and
spontaneous internal fluctuations is not settled \cite{ruelle}.
In many complex systems, this question amounts to
distinguishing between endogeneity and exogeneity and is important
for understanding the relative effects of self-organization versus
external impacts. This is difficult in most physical systems because
externally imposed perturbations may lie outside the complex attractor
which itself may exhibit bifurcations. Therefore, observable perturbations
are often misclassified. 

It is thus interesting to study other systems, in which the dividing line
between endogenous and exogenous shocks may be clearer in the hope that it
would lead to insight about complex physical systems. The investigations
of the two questions above may also bring new understanding of these systems.
The systems to which the endogenous-exogenous question (which we will refer to as ``endo-exo''
for short) is relevant include the following:
\begin{itemize}
\item Biological extinctions such as the
Cretaceous/Tertiary KT boundary (meteorite versus extreme volcanic
activity (Deccan traps) versus self-organized critical extinction cascades), 

\item immune system deficiencies (external viral/bacterial infections versus
internal cascades of regulatory breakdowns), 

\item cognition and brain learning processes (role of
external inputs versus internal self-organization and
reinforcements),

\item discoveries (serendipity versus the outcome of slow endogenous maturation
processes), 

\item commercial successes (progressive reputation cascade versus the result of a well
orchestrated advertisement),

\item financial crashes (external shocks versus self-organized instability),

\item intermittent bursts of financial volatility (external shocks versus
cumulative effects of news in a long-memory system),

\item the aviation industry
recession (9/11/2001 terrorist attack versus structural endogenous problems),

\item social unrests (triggering factor or rotting of social tissue),

\item recovery after wars (internally generated
(civil wars) versus imported from the outside) and so on.
\end{itemize}

It is interesting to mention that the question of exogenous versus endogenous forcing has been
hotly debated in economics for decades. A prominent example is the
theory of Schumpeter on the importance of technological discontinuities in economic history.
Schumpeter argued that ``evolution is
lopsided, discontinuous, disharmonious by nature... studded with violent outbursts and
catastrophes... more like a series of explosions than a gentle, though incessant,
transformation'' \cite{Schumpeter}. Endogeneity versus exogeneity is also
paramount in economic growth theory \cite{Romer}.
Our analyses reviewed below suggest a subtle interplay between exogenous and endogenous shocks
which may cast a new light on this debate.

In the sequel, we review the works of the author with his collaborators, which
have investigated the endo-exo question in a variety of systems.

\section{Exogenous and endogenous shocks in social networks}

One defining characteristics of humans is their organization in
social networks. It is probable that the large brain, which makes what
we are, has been shaped by social interactions, and may have co-evolved
with the size and complexity of social groups \cite{SBH,ZhouDunbar}.
A single individual may belong to several intertwinned social networks, associated with their
different activities (work colleagues, college alumny societies, friends,
family members, etc.). The formation and the evolution of social networks
and their mutual entanglements control the hierarchy of interactions between
humans, from the individual level to society and to culture. In this section,
we review a few original probes of several social networks which unearth
a remarkable universality: the distribution of human decision times in social networks
seem to be described by a power law $1/t^{1+\theta}$ with $\theta =0.3 \pm 0.1$.
This constitutes an essential ingredient in models describing how the cascade of agent decisions
leads to the bottom-up organization of the response of social systems. 
We first present such a model in terms of a simple epidemic process of word-of-mouth effects
\cite{ETAS,SHendo,Amazon_paper} and then discuss the different data sets.

\subsection{A simple epidemic cascade model of social interactions \label{theo}}

Let us consider an observable characterizing the activity of humans
within a given social network of interactions. This activity can 
be the rate of visits or downloads on an internet website, the sales of a book
or the number of newpaper articles on a given subject. 

We envision that the instantaneous activity
results from a combination of external forces such as
news and advertisement, and of social influences in
which each past active individual may impregnate other individuals in
her network of acquaintances with the desire to act. This
impact of an active individual onto other humans is not instantaneous as
people react at a variety of time scales. The time delays capture
the time interval between social encounters, the maturation of 
the decision process which can be influenced by mood, sentiments, and many
other factors and the availabilty and capacity to implement the decision.
We postulate that this latency can be
described by a memory kernel $\phi(t-t_i)$ giving the probability
that an action at time $t_i$ leads to another action at a later time $t$
by another person in direct contact with the first active individual. 
We consider the memory function $\phi(t-t_i)$ as a fundamental macroscopic description
of how long it takes for a human to be triggered into action, following
the interaction with an already active human.

Then, starting
from an initial active individual (the ``mother'') who first acts
(either from exogenous news or by chance), she may trigger action
by first-generation ``daughters,'' which themselves
propagate the drive to act to their own friends, who become
second-generation active individuals, and so on. This cascade of generations
can be shown to renormalize the memory kernel $\phi(t-t_i)$ into a
dressed or renormalized memory kernel $K(t-t_i)$
\cite{SS99,ETAS,SHendo}, giving the probability that an action at time
$t_i$ leads to another action by another person at a later time $t$
through any possible generation lineage. In physical terminology,
the renormalized memory kernel $K(t)$ is nothing but the response
function of the system to an impulse. This is captured by the following equations
\be
A(t) = s(t) + \int_{-\infty}^t d\tau ~ A(\tau)~\phi(t-\tau)
= \int_{-\infty}^t d\tau ~ s(\tau)~K(t-\tau)~.
\label{gmmls}
\ee
The meaning of these two equivalent formulations is the following. The $s(t)$'s are the 
spontaneous exogenous activations. The integral $\int_{-\infty}^t d\tau ~ A(\tau)~\phi(t-\tau)$
gives the additional contribution due to past activities $A(\tau)$ whose influences to the present
are mediated by the direct influence kernel $\phi$ of first generation. The last integral 
$\int_{-\infty}^t d\tau ~ s(\tau)~K(t-\tau)$ expresses the fact that the present activity
$A(t)$ can also be seen as resulting from all past exogenous sources $s(\tau)$ mediated to the present
by the renormalized kernel $K$, which takes into account all generations of 
cascades of influences.

The following functional dependence
is found to provide an accurate description, as we shall discuss below:
\be 
K(t) \sim 1/(t-t_c)^{p}~, ~~~~{\rm with}~~p=1-\theta~.
\label{gmlasa} 
\ee 
The dependence (\ref{gmlasa}) implies \cite{SS99,ETAS,SHendo}
\be
\phi(t) \sim 1/(t-t_c)^{1+\theta}~.
\label{mgmdfl}
\ee

We should stress that the renormalization from the usually (but not always)
unobservable ``bare'' response function $\phi(t)$
with exponent $1+\theta$
in (\ref{mgmdfl}) to the observable ``renormalized'' response function $K(t)$
in (\ref{gmlasa}) with exponent $1-\theta$ is obtained if the network
is close to critical, i.e., if the average branching ratio $n$ is close to $1$ ($n$ is 
defined as the average number of daughters of first generation per mother). In other
words, there is on average approximately one triggered daughter per active mother.
This condition of criticality ensures, in the language of branching processes, 
that avalanches of active people triggered by a given mother are self-similar (power
law distributed). In contrast, for $n<1$, the cascade of triggered actions
is ``sub-critical'' and avalanches die off more rapidly. It can be shown
\cite{SS99,ETAS,SHendo} that there is in this case a characteristic time scale
\be
t^* \sim {1 \over (1-n)^{1/\theta}}~
\label{mgks}
\ee
acting like a correlation time, which separates two regimes:
\begin{itemize}
\item for $t<t^*$, the renormalized response function $K(t)$ is indeed of the form
(\ref{gmlasa});
\item for $t>t^*$, the renormalized response function $K(t)$ crosses over to 
an asymptotic decay with exponent $1+\theta$, of the form of $\phi(t)$ in (\ref{mgmdfl}).
\end{itemize}
For $n>1$, the epidemic process is super-critical and has a finite probability
of growing exponentially fast. We will not be concerned with this last regime which does
not seem relevant in the data discussed below.

In the absence of strong external influences, a peak in social activity
can occur spontaneously due to the interplay between a continuous
stochastic flow of small external news and the amplifying impact
of the epidemic cascade of social influences.  It can then be shown that,
for $n$ close to $1$ or equivalently for $|t-t_c|<t^*$, the average
growth of the social activity before such an ``endogenous'' peak and the
relaxation after the peak are proportional to
\cite{SHendo,HSG}
\be 
\int_0^{+\infty}K(t-t_c+u)K(u)du
\sim 1/|t-t_c|^{1-2\theta}~, 
\label{mvmlw} 
\ee 
where the
right-hand-side expression holds for $K(t)$ of the form
(\ref{gmlasa}). The prediction that the relaxation following an
exogeneous shock should happen faster (larger exponent $1-\theta$)
than for an endogeneous shock (with exponent $1-2\theta$) 
agrees with the intuition that an endogeneous shock should have
inpregnated the network much more and thus have
a longer lived influence. In a nutshell,
the mechanism at the origin of the endogenous response function
(\ref{mvmlw}) is the constructive interference of accumulated
small news cascading through the social influence network. In
other words, the presence of a hierarchy of nested relaxations
$K(t)$ given by (\ref{gmlasa}), each one associated with each
small news, creates the effective endogenous response (\ref{mvmlw}).

Dodds and Watts have recently introduce a general model of contagion which, by explicitly incorporating
memory of past exposures to, for example, an infectious agent, rumor, or new
product, includes the main features of existing contagion models and
interpolates between them \cite{Dodds}.

\subsection{Internet download shocks}

\begin{figure}
\begin{center}
\epsfig{file=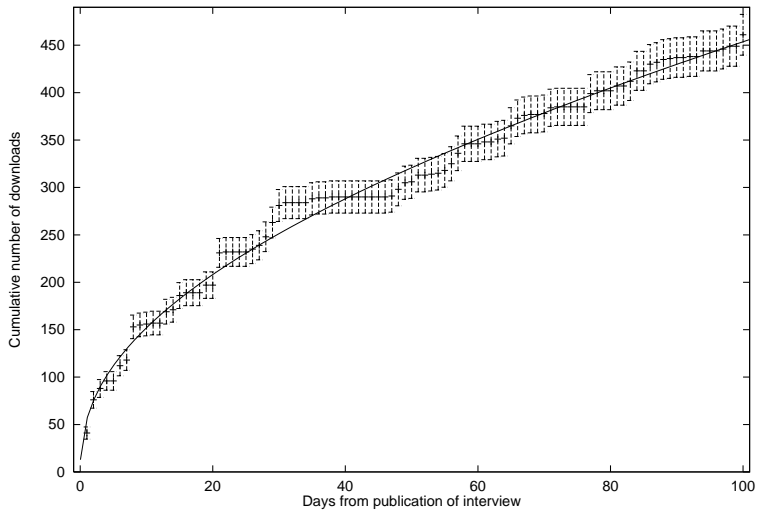,width=10cm}
\caption{\protect\label{hitsfit} Cumulative number of downloads $N$ as a
function of time $t$ from the appearance of the interview on Wednesday
the 14 April 1999. The fit is $N(t) = \frac{a}{1-p} t^{1-p} + ct$ with $b
\approx 0.58 \pm 0.03$. Reproduced from \cite{JSdownload1}.}
\end{center}
\end{figure}

In Ref.~\cite{JSdownload1}, Johansen and Sornette report the following
experiment. The authors were interviewed by a journalist
from the leading Danish newspaper JyllandsPosten on a subject of rather
broad and catchy interest, namely stock market crashes. The interview was
published on April 14, 1999 in both the paper version of the newspaper
as well as in the electronic version (with access restricted to subscribers)
and included the URLs where the authors' research papers on the subject could be retrieved.
It was hence possible to monitor the number of downloads of papers 
as a function of time since the publication date of the interview.
The rate of downloads of the authors' papers as a function of time was
found to obey a $1/t^p$ power law with exponent $b=0.58\pm 0.03$, as shown
in figure \ref{hitsfit}.

Within the model of epidemic word-of-mouth effect summarized in
section \ref{theo}, the relaxation of the rate
of downloads after the publication of the interview characterizes the response
function $K(t)$ given by (\ref{gmlasa}) with respect to
an exogenous peak: prior to the publication of the interview, the rate of 
downloads was slightly less than one per day; it suddenly jumped to several
tens of downloads per day in the first few days after 
the publication and then relaxed slowly according to (\ref{gmlasa}).
The reported power law with exponent $p \simeq 0.6$ is compatible with the form 
(\ref{gmlasa}) with $\theta=0.4$, which is within the range of 
other values: $\theta =0.3 \pm 0.1$.

Johansen \cite{J2web} has reported another similar observation following another
web-interview on stock market crashes, which contained the URL of his
articles on the subject. He found again a power law dependence
(\ref{gmlasa}), but with an exponent $p$ close to $1$, leading 
in the terminology of the model of epidemic word-of-mouth effect to $\theta \simeq 0$.
Two interpretations are possible: (i) the exponent $\theta$ is non-universal; (ii)
the social network is not always close to criticality ($n \simeq 1$) and the 
observable response function $K(t)$ is then expected to cross-over smoothly
from a power law with exponent $1-\theta$ to another asymptotic power law with
exponent $1+\theta$. According to this second hypothesis, the exponent $p$
of the relaxation kernel $K(t)$ may be found in the range $1-\theta$ to $1+\theta$,
depending upon the range of investigated time scales and the proximity $1-n$ to
criticality. We find hypothesis (ii) more attractive
as it puts the blame on the non--universal parameter $n$, which embodies the 
connectivity structure, static and dynamics, of social interactions at a given moment.
It does not seem unrealistic to think that $n$ may not be always at its 
critical value $1$, due to many possible other social influences. In constrast,
one could postulate that the power law (\ref{mgmdfl}) for the direct influence function
$\phi(t)$ between two directly linked humans may reflect a more universal character.
But, of course, only more empirical investigations will allow to put
more light on this issue.

Eckmann, Moses and Sergi \cite{Eck} also report an original investigation
probing the temporal dynamics of social networks using email networks
in their universities. They find a distribution
of response times till a message is answered which seems to be a power
law with exponent less than $1$ at early times (1 hour) to another power law with
exponent larger than $1$ at long times (days), which could be a direct evidence
of the direct response function $\phi(t)$ defined in (\ref{mgmdfl}).
The relationship between their investigation and the
previous works using web downloads \cite{JSdownload1,J2web} has been
noted by Johansen \cite{notedja}.

\subsection{Book sale shocks}

\begin{figure}
\begin{center}
         \epsfig{file=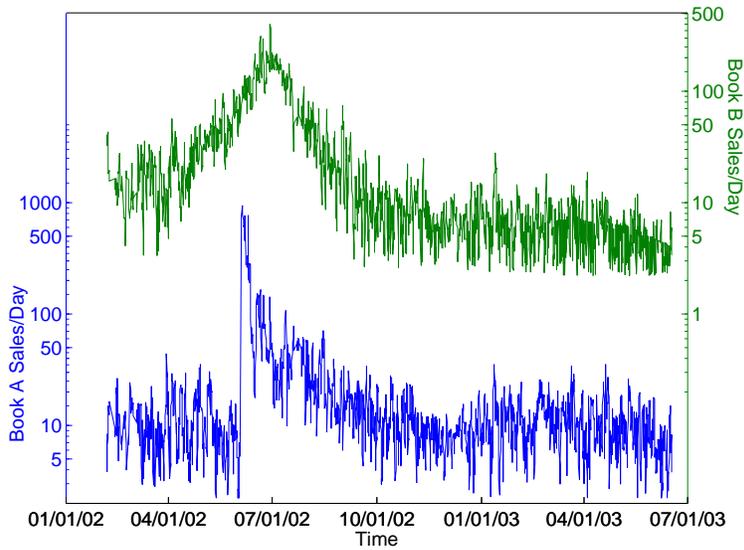,width=10cm}
     \caption{\label{fig2} Time evolution over a year and a half of
    the sales per day of two books: Book A (bottom, blue, left scale) 
    is ``Strong Women Stay Young'' by Dr. M.
    Nelson and Book B (top, green, right scale) is ``Heaven and Earth 
    (Three Sisters Island Trilogy)'' by N. Roberts.
    The difference in the patterns is striking, Book A undergoing an exogenous peak on June 5, 2002,
    and Book B endogenously reaching a maximum on June 29, 2002. 
    Reproduced from \protect\cite{Amazon_paper}.}
     \end{center}
\end{figure}

Sornette, Deschatres, Gilbert and Ageon have used a database of sales 
from Amazon.com as a proxy for commercial growth and successes \cite{Amazon_paper}.
Fig. \ref{fig2} shows about 1.5 years of data for two
books, Book A (``Strong Women Stay Young'' by Dr. M. Nelson) and
Book B (``Heaven and Earth (Three Sisters Island Trilogy)'' by N.
Roberts), which are illustrative of the two classes found in this
study. Book A jumped on June 5, 2002, from rank in the 2,000s to
rank 6 in less than 12 hours. On June 4, 2002, the New York Times
published an article crediting the ``groundbreaking research done
by Dr. Miriam Nelson'' and advising the female reader, interested
in having a youthful postmenopausal body, to buy the book and
consult it directly \cite{NYT}. This case is the archetype of an
``exogenous'' shock. In contrast, Book B culminated at the end of
June 2002 after a slow and continuous growth, with no such
newspaper article, followed by a similar almost symmetrical decay,
the entire process taking about 4 months. We will show below that
the peak for Book B belongs to the class of endogenous shocks.
Qualitatively, such endogeneous growth is well explained in Ref.
\cite{tipping} by taking the example of the book ``Divine Secrets
of the Ya-Ya Sisterhood'' by R. Wells, which became a bestseller
two years after publication, with no major advertising campaign.
Following the reading of this originally small budget book,
``Women began forming \textit{Ya-Ya} Sisterhood groups of their
own [...]. The word about \textit{Ya-Ya} was spreading [...] from
reading group to reading group, from \textit{Ya-Ya} group to
\textit{Ya-Ya} group'' \cite{tipping}. Generally, the popularity
of a book is based on whether the information regarding that book
will be able to propagate far and long enough into the network of
potential buyers.

\begin{figure}
   \begin{center}
    \epsfig{file=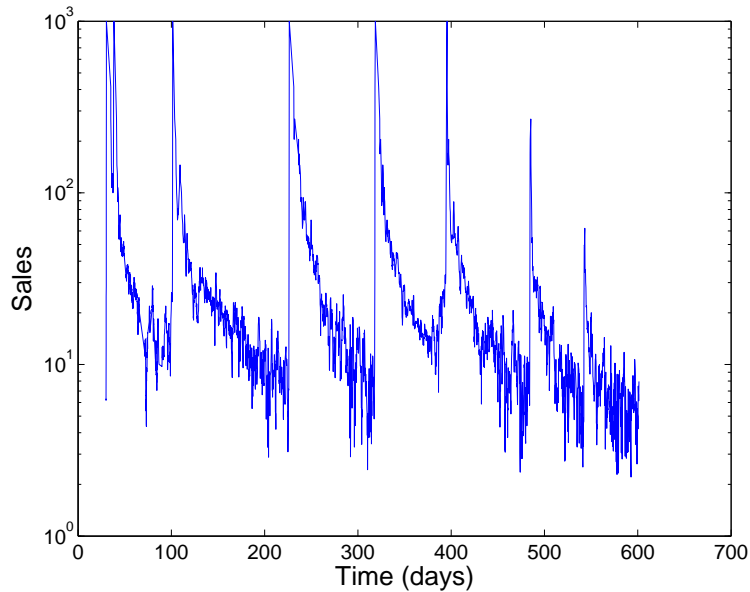,width=10cm}
     \caption{\label{oprah} Time evolution of the book entitled ``Get with the Program.'' 
     Each time the book appeared on Oprah Winfrey Show
     (B.Greene is O. Winfrey's trainer), the sales jumped overnight.}
   \end{center}
\end{figure}

Another dramatic example of exogenous shocks is shown in figure \ref{oprah}:
here the personal trainer of Oprah Winfrey had his book presented $7$ or $8$ times
during the Oprah Winfrey Show, leading to dramatic overnight jumps in sales.

\begin{figure}
     \begin{center}
         \epsfig{file=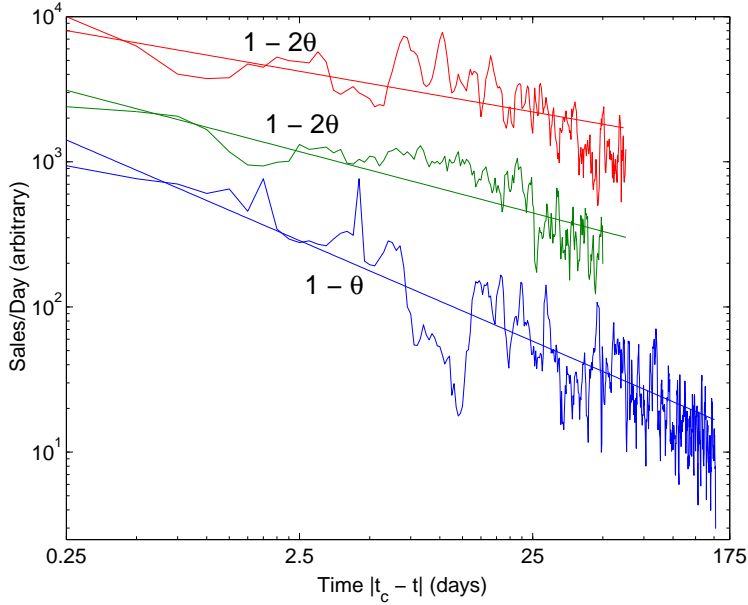, width=10cm}
     \caption{\label{fig3} The bottom curve (blue) shows the
     relaxation of the sales of Book A after the peak of $t_c =$ June 5, 2002 as a function
     of the time $t-t_c$ from the time of the peak. The least
     squares best fit with a power law gives a slope $\approx -0.7$. Since this
     peak is identified as exogenous with theoretical slope $1-\theta$,
     we get the estimate $\theta = 0.3 \pm 0.1$. The curve
     in the middle (green shifted up by a factor $6$) shows the relaxation of sales of Book B
     after the peak of $t_c=$ June 29, 2002 as a function
     of the time $t-t_c$ from the time of the peak. The least squares fit gives a
     slope of $\approx -0.4$, which provides the 
     independent estimate $\theta = 0.3 \pm 0.1$ from the
     theoretical endogenous exponent $1-2\theta$.  The top curve (red
     shifted up by a factor 25 with respect to the bottom curve) shows
     the acceleration of the sales of Book B leading to the same peak
     at $t_c=$ June 29, 2002 as a function of the time $t_c-t$ to the time of the peak.
     The time
     on the $x$-axis has been reversed to compare the precursory acceleration
     with the aftershock relaxation. The least squares slope is $\approx -0.3$
     not far from the prediction $1-2\theta$ of the cascade model, with $\theta = 0.3 \pm 0.1$.}
      \end{center}
\end{figure}

Each relaxation of sales for about 140 books that reached the top 50
in the Amazon.com ranking system have been analyzed and
shown to fall into two categories: the relaxations described by a power law with
an exponent close to $0.7 = 1-\theta$ and the relaxations described by a power
law with an exponent close to $0.4=1-2 \theta$, for $\theta \simeq 0.3$.  An example
of such fits for the two books shown in figure \ref{fig2} is presented in figure \ref{fig3}.
In addition, Sornette et al. \cite{Amazon_paper} checked that an overwhelming majority
of those sale peaks classified as exogenous from the value of their exponent 
$\simeq 0.7 = 1-\theta$ were preceded by an abrupt jump, in agreement with the 
epidemic cascade model of social interactions described in section \ref{theo}.
In constrast, those sale peaks falling in the endogenous class according to their
exponent $\simeq 0.4=1-2 \theta$ of their relaxation after the peak were found to 
be preceded by an approximately symmetry growth described by a power law with the
same exponent, as predicted by (\ref{mvmlw}). An example is shown also for book B
in figure \ref{fig3}.

The small values of the exponents (close to $1-\theta$ and $1-2\theta$)
both for exogenous and endogenous relaxations
imply that the sales dynamics is dominated by cascades involving 
high-order generations rather than by interactions stopping after first-generation 
buy triggering. Indeed, if buys were initiated mostly by the 
direct effects of news or advertisements, and not much by triggering cascades
in the acquaintance network, the cascade model predicts that we should then measure an
exponent $1+\theta$ given by the ``bare'' memory kernel $\phi(t)$, as
already said.
This implies that the average number $n$ 
(the average branching ratio in the language of branching models)
of impregnated buyers per initial
buyer in the social epidemic model is on average very close to 
the critical value $1$, because the renormalization from $\phi(t)$ to
$K(t)$ given by (\ref{gmlasa}) only operates close to criticality
characterized by the occurrence of large cascades of buys.
Reciprocally, a value of the exponent $p$ larger than $1$ suggests
that the associated social network is far from critical. Such
instances can actually be observed. Examples of cross-overs 
from the renormalized response function $K(t)$ (\ref{gmlasa}) to $\phi(t)$ in (\ref{mgmdfl})
with an asymptotic decay with exponent $1+\theta$ has been 
documented \cite{Amazon_paper,Amazon_paper2}.
Note that it is possible to give an
analytical description of this cross-over exhibited by $K(t)$ as a
function $n$ \cite{ETAS}, thus allowing in principle to invert for
$n$ for a given data set. This opens the tantalizing possibility of measuring the
dynamical connectivity of the social network, and possibly to
monitor it as a function of time. 

This findings open other interesting avenues of research. While 
this first investigation has emphasized the distinction
between exogenous and endogenous peaks to set the fundamentals for
a general study, repeating peaks as well as peaks
that may not be pure members of a single class are also frequent. 
In a sense, there are no real ``endogenous'' peaks, one could argue, because
there is always a source or a string of news impacting
on the network of buyers. What Sornette et al. \cite{Amazon_paper} have done is to
distinguish between two extremes, the very large news impact and
the structureless flow of small news amplified by the cascade effect
within the network. One can imagine and actually observe a continuum
between these two extremes, with feedbacks between the development of
endogeneous peaks and the attraction of interest of the media as a
consequence, feeding back and providing a kind of exogenous boost, and
so on.  In those and in more
complicated cases, the epidemic model of word-of-mouth effects 
should provide a starting platform to predict the sales
dynamics as a function of an arbitrary set of external sources. 
Tracking dynamically the connectivity $n(t)$ of each social network
relevant to a given product, it should also be possible to target
the most favorable times, corresponding to the largest $n(t)$, for promoting or
sustaining the sales of a given product, with obvious consequences
for marketing and advertisement strategies. An additional extension
includes the possible feedback of the marketing strategy on the
control parameter $n(t)$ which could be manipulated so as to 
keep the system critical, an ideal situation from the point of view
of marketers and firms. Quantifying this effect requires to extend
the simple epidemic model in the spirit of mechanisms leading
to self-organized criticality by positive feedbacks of the order
parameter onto the control parameter \cite{feedsoc,gilsor}. Sornette et al.'s results 
suggest that social networks have evolved to converge very close
to criticality. As Andreas S. Weigend,
chief scientist of Amazon.com (2002-2004) wrote on his webpage: ``Amazon.com might be 
the world's largest laboratory to study human
behavior and decision making.''  I share this view point.

Actually, I envision that an extension of Sornette et al.'s study to
a broad database of sales from all products sold by e-retailers like Amazon.com
could give access to the equivalent of the ``social climate'' of a country
such as the USA and its evolution as a function of time under
the various exogenous and endogenous factors at work. 
Indeed, Amazon.com categorizes its products in different
(tradable) dimensions of possible interest to a human being, such as
\begin{itemize}
\item Books, Music, DVD,
\item Electronics (audia and video, camera and photo, software, computer and video games, cell phones...),
\item Office,
\item Kids and Baby,
\item Home and Garden (which includes pets),
\item Gifts, Registries, Jewelry and watches,
\item Apparel and Accessories,
\item Food,
\item Health, Personal Care, Beauty,
\item Sports and Outdoors,
\item Services (movies, restaurants, travel, cars, ...),
\item Arts and Hobbies,
\item Friend and Favorites,
\end{itemize}
with many sub-categories. Monitoring and analyzing the sales as a function of
time in these different categories is like getting the temperature, wind velocity, 
humidity in meteorology in many different locations. The flow of interest of society at large and of
sub-groups could in principle tells us how society is responding in its
spending habits to large scale influence. 
As an illustrative example, it has
been shown that, during bullish periods characterized by strong stock market gains
(bubble regimes), the number of books written and sold related to financial
investments soared \cite{RSfrenzy,Roehnerspec}.

Another potentially fruitful application is the music industry and the impact
on sales versus Internet piracy
of the quality of performers (endogenous effect on the network of potential 
buyers who can promote a CD by word-and-mouth in the network of 
potential buyers) versus the promotion campaigns
of short-lived performers and their one-only-hit wonders \cite{music}.
Indeed, according to an internal study done by one of the big companies that dominate
the production and distribution of music, 
the drop in sales in America may have less to do with internet piracy than with
other factors, among them the decreasing quality of music itself.
The days of watching a band develop slowly over time with live
performances are over, according to some professionals. Even Wall Street analysts are questioning
quality. If CD sales have shrunk, one reason could be that people are less excited by
the industry's product. A poll by Rolling Stone magazine found that fans, at least,
believe that relatively few ``great'' albums have been produced recently \cite{music}.
This is clearly an endo-exo question that can be analyzed with databases available
on the Internet.

\subsection{Social shocks}

Roehner, Sornette and Andersen \cite{RSA} have used the concept of exogeneous shocks
to propose a general method for quantifying the response
function in order to advance the social sciences. 
By using a database of newspaper
articles called Lexis-Nexis, which is available in many
departments of political science or sociology, they have quantified 
the response to shocks, such as the following:
\begin{itemize}
\item On October, 31, 1984, the
Prime minister of India, Indira Gandhi, was assassinated
by two of her Sikh bodyguards. This event triggered a
wave of retaliations against Sikh people and Sikh property, not only
in India (particularly in New Delhi), but in many other
countries as well. 

\item In the early hours of December, 6, 1992, thousands of Hindus converged
toward the holy city of Ayodhya in northern India and began
to destroy the Babri mosque which was said to be built on the
birthplace of Lord Rama. The old brick walls came down fairly
easily and soon the three domes of the mosque crashed to the
ground. This event triggered a burst of protestations and
retaliations which swept the whole world from Bangladesh to
Pakistan, to England or the Netherlands. In all these
countries, Hindu people
were assaulted, Hindu temples were firebombed, damaged or destroyed.

\item On September, 11, 2001, two planes
crashed into the twin towers of the World Trade Center in
New York. This event triggered a wave of reactions
against Islamic people and property
not only in the United States but also in other countries
\end{itemize}

\begin{figure}
\begin{center}
         \epsfig{file=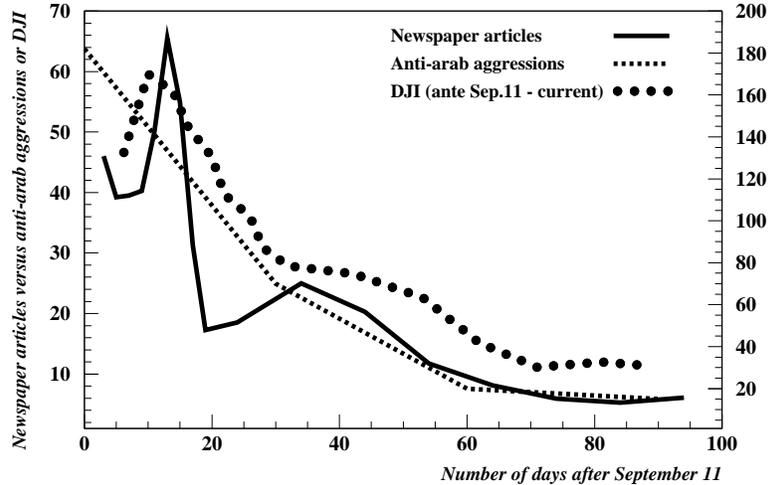,width=10cm}
     \caption{\label{figqua2} Relaxation of three different
     social variables after the shock of September 11, 2001.
The solid line curve is the number of articles writing on the destruction of
mosques after the event; the broken line (scale on the
right-hand side) shows the number of anti-arab aggressions
in California in the three months after September 11;
the dotted line shows the changes in the level of the
Dow Jones Index with respect to its pre-Sep.11 level as given by
the difference DJI(pre-9/11)-DJI(current).
Source: California's Attorney General Office,
published in the San Jose Mercury News, 11 March 2002.
Reproduced from \protect\cite{RSA}.}
     \end{center}
\end{figure}

For these different events, Roehner et al. \cite{RSA} show that different quantitative measures of 
social responses exhibit an approximate universal behavior, again characterized
by a power law, as shown in figure \ref{figqua2}. This figure gives the 
time evolution after September 11, 2001 of newspaper articles, anti-arab agressions
and the Dow Jones Industrial Average, which are approximated by a power law
$\sim 1/t^{p}$. Due to the coarseness of the measures, the exponent $p$ is
not well-constrained: $p=-1.8 \pm 0.7$ (newspaper articles),
$p=-1.4 \pm 0.5$ (anti-arab agressions) and $p=-2.2\pm 1.6 $ (DJI).
Comparing the reaction to September 11, 2001 in different countries such 
as Canada, Great Britain and the Netherlands, Roehner et al. \cite{RSA} have
suggested that the response function actually expresses an information 
on ``cracks'' pre-existing in the social networks of the corresponding countries.
For instance, the number of attacks to Mosques has been larger in the Netherlands, 
which is in line with other information on the 
concerns at high political levels (private communication to the authors) about the integrity of
the social tissue in the Netherlands, a fact illustrated
more recently on the political scene by the rapid rise and then
assassination of the rightist politician Fortuyn in May 2002.
This line of evidence could be quantified within the epidemic model of social influence
by different values of the connectivity parameter $n$ in different countries.

Burch, Emery and Fuerst \cite{Burch} have used also the unique
opportunity offered by the ``nine-eleven'' 
terrorist attack to confirm clearly the hypothesis that
closed-end mutual fund discounts from fund net asset values reflect small investor
sentiment. Carter and Simkins \cite{Simkins} investigated the reaction of airline stock prices
 to the 9/11 terrorist attack and found that the market was concerned
 about the increased likelihood of bankruptcy in the wake of the
 attacks and distinguished between airlines based on their
 ability to cover short-term obligations (i.e., liquidity).

\section{Exogenous and endogenous shocks in financial markets}

\subsection{Volatility shocks}

Standard economic theory holds that the complex trajectory of stock
market prices is the faithful reflection of the continuous flow of news
that are interpreted and digested by an army of analysts and traders.
Accordingly, large shocks should result from really bad surprises. It is
a fact that exogenous shocks exist, as epitomized by the recent events
of Sept. 11, 2001, and there is no doubt about the existence of utterly
exogenous bad news that move stock market prices and create strong
bursts of volatility. A case that cannot be refuted is the the market
turmoil in Japan following the Kobe earthquake of Jan. 17, 1995 that led
to a total cost estimated around \$200 billion dollars. Indeed, as long
as the science of earthquake prediction is still in its infancy,
destructive earthquakes are not endogeneized in advance in stock market
prices by rational agents ignorant of seismological processes. One may
also argue that the invasion of Koweit by Iraq on Aug. 2, 1990 and the
coup against Gorbachev on Aug., 19, 1991 were strong exogenous shocks.
However, some could argue that precursory fingerprints of these events
were known to some insiders, suggesting the possibility that the action of
these informed agents may have been reflected in part in stock markets
prices. Even more difficult is the classification (endogenous versus
exogenous) of the hierarchy of volatility bursts that continuously shake
stock markets. While it is a common practice to associate the large
market moves and strong bursts of volatility with external economic,
political or natural events \cite{white}, there is not convincing
evidence supporting it.

Perhaps the most robust observation
in financial stock markets is that volatility is serially correlated with
long-term dependence (approximately power law like). Volatility autocorrelation
is typically modeled using autoregressive conditional heteroskedasticity (ARCH)
\cite{Engle}, generalized ARCH \cite{Bollerslev}, stochastic volatility \cite{Anderson},
Markov switching \cite{Hamilton,Hamilton2}, nonparametric \cite{Pagan} and extensions
of these models (see \cite{sche} for comparisons). Recent 
powerful extensions include the Multifractal Random
Walk model (MRW) introduced by Muzy, Bacri and Delour \cite{M_etal,B_etal}, which belongs
to the class of stochastic volatility models. Using the MRW, 
Sornette, Malevergne and Muzy \cite{VolMRW} have shown that it is possible
to distinguish between an endogenous and an exogenous origin to a volatility shock.
Tests on the Oct. 1987 crash, on
a hierarchy of volatility shocks and on a few of the obvious exogenous shocks have
validated the concept.
This study shows that the relaxation with time of a burst of volatility is distinctly
different after a strong exogenous shock compared with the relaxation of 
volatility after a peak with no identifiable 
exogenous sources. This study does not explain the origin of volatility correlation.
But it identifies the ``natural'' response function of the system to an 
external shock, from which
the stationary long-term dependence structure of the volatility and its
intermittent bursts derive automatically. In other words, Sornette et al.'s
study leads to the view that the properties of the volatility can be 
in large part understood from a single characteristic, which is the response
of the agents to a new piece of news. This response function must
ultimately be derived from the behavior of financial agents, for instance
taking into account their sensitivity to wealth changes, their loss aversion
as well as their finite-time memory of past losses that may impact their
future decision \cite{queen}.

The multifractal random walk is an autoregressive process
with a long-range memory decaying as $t^{-1/2}$, which is defined
on the logarithm of the volatility. Using the MRW model for the dependence
structure of the volatility, Sornette et al.  predict that 
exogenous volatility shocks will be followed by a universal relaxation
\be
\simeq \lambda/t^{1/2}~,
\label{g,dd}
\ee 
where $\lambda$ is the multifractal parameter,
while endogenous volatility shocks relax according
to a power law
\be
\simeq 1/t^{p(V_0)}~,~~~~~{\rm with}~~p(V_0) \simeq \lambda^2~ \ln(V_0)~,
\label{vmkmele}
\ee
with an exponent $p(V_0)$ which is a linear function of 
the logarithm $\ln(V_0)$ of the shock of volatility $V_0$.
The difference between these behaviors and those reported above
modeled by the epidemic process with long-term memory stems from the fact that 
the stock market returns $r_{\Dt}(t)$ at time scale $\Dt$ at 
a given time $t$ can be accurately described by the following process \cite{M_etal,B_etal}:
\be
r_{\Dt}(t) = \epsilon(t) \cdot  \sigma_{\Dt}(t)  = \epsilon(t)
\cdot e^{\w_{\Dt}(t)}~,
\label{remglww}
\ee
where $\epsilon(t)$ is a standardized Gaussian 
white noise independent of $\w_{\Dt}(t)$ and
$\w_{\Dt}(t)$ is a nearly Gaussian process with mean and
covariance:
\ba
\label{muo}
\mu_{\Dt} & = & {1 \over 2} \ln(\sigma^2 \Dt)-C_\Dt(0) \\
C_\Dt(\tau) & = & \Cov[\w_{\Dt}(t), \w_{\Dt}(t + \tau)] =
\lambda^2 \ln \left( \frac{T}{|\tau|+e^{-3/2}\Dt} \right)~.
\label{vo}
\ea
where $\sigma^2 \Dt$ is the return variance at scale $\Dt$ and 
$T$ represents an ``integral'' (correlation) 
time scale. $\lambda$ is called the multifractal parameter: when it 
vanishes, the MRW reduces to a standard Wiener process (standard continuous
random walk).
Such logarithmic decay of log-volatility covariance
at different time scales has been evidenced empirically
in \cite{B_etal,M_etal}. Typical values for $T$ and $\lambda^2$ are
respectively $1$ year and $0.04$. 

The MRW model can be expressed in a more familiar form, in which the
log-volatility $\omega_\Dt(t)$ obeys an auto-regressive equation whose 
solution reads
\be
\omega_\Dt(t) = \mu_\Dt+\int_{-\infty}^t d\tau~ \eta(\tau)~K_\Dt(t-\tau)~,
\label{mbhmle}
\ee
where  $\eta(t)$ denotes a
standardized 
Gaussian white noise and
the memory kernel $K_\Dt(\cdot)$ is a causal function, ensuring that the 
system is not
anticipative.
The process $\eta(t)$ can be seen as the information flow. Thus
$\w(t)$ represents the response of the market to incoming information up to the
date $t$. At time $t$, the distribution of $\w_\Dt(t)$ is Gaussian with mean
$\mu_\Dt$ and variance
$V_\Dt = \int_0 ^\infty d\tau~
K^2_\Dt(\tau) = \lambda^2 \ln \left( \frac{Te^{3/2}}{\Delta t} \right)$.
Its covariance, which entirely specifies the random process, is given by
\be
C_\Dt(\tau) = \int_0 ^\infty dt~ K_\Dt(t) K_\Dt(t+|\tau|)~.
\label{eq:ker}
\ee
Performing a Fourier transform, we obtain
\be
\hat K_\Dt(f)^2 = \hat C_\Dt(f) = 2 \lambda^2~ f^{-1}\left[\int_0^{Tf}{\sin(t) \over
    t} dt+O\left(f\Dt\ln(f\Dt)\right)\right]~,
\ee
which shows, using (\ref{vo}), that for $\tau$ small enough,
\be
K_\Dt(\tau) \sim K_0 \sqrt{\frac{\lambda^2 T}{\tau}} ~~~~~ \mbox{for}~~ \Dt << \tau << T~,
\label{mgmlww}
\ee
which is the above stated exogenous response function (\ref{g,dd}).
This slow power law decay (\ref{mgmlww}) of the memory kernel in (\ref{mbhmle})
ensures the long-range dependence and multifractality of the 
stochastic volatility
process (\ref{remglww}). 

The main difference between the MRW model and the previous class of epidemic process
is that the long-term memory appear in the logarithm of the variable in the former, 
as shown from equation (\ref{mbhmle}). As a consequence, the MRW basically 
describes a variable with is the exponential of a long-memory process. It is 
the interplay between this strongly nonlinear exponentiation and the long-memory
which gives the multifractal properties to the MRW and, as a consequence, the
shock amplitude dependence of the exponents $p(r)$ of the relaxation of the volatility
following endogenous shocks. In contrast, the linear long-term memory structure (\ref{gmmls})
of the epidemic processes of section \ref{theo} ensures universal exponents
which are independent of the shock amplitudes (but not of the endo-versus-exo nature).
In the epidemic process (\ref{gmmls}), the relationship between exogenous and 
endogenous relaxations is expressed by the exponents of the power laws
$\sim 1/t^{1-\theta}$ (exo) versus $\sim 1/t^{1-2\theta}$ (endo). In the MRW, notice that
the relationship between exogenous (\ref{g,dd}) and endogenous relaxations 
(\ref{vmkmele}) is through the multifractal parameter $\lambda$: the fact that 
an amplitude of the exogenous response function impacts the power law exponent
of the endogenous relaxation is again a signature of the exponential structure
of the multifractal model. The MRW extends the realm of possible relationships
between endogenous and exogenous responses discussed until now.

\subsection{Financial crashes}

The endo-exo question also appears to be crucial for understanding financial crashes.
In contrast with the previous examples, the distinction is not that much
in the relaxation or recovery after the shock but rather in the precursory
behavior before the crash. An endogenous crash might be expected 
to end a period of strong price gains due for instance to speculative herding. In
contrast, an exogenous crash would be the response of the financial system
to a very strong adverse piece of information.

Indeed, according to standard economic theory, the complex 
trajectory of stock market prices is the faithful reflection of the continuous 
flow of news that are interpreted and digested by an army of analysts and 
traders \cite{Cutler}. Accordingly, large market losses should result from 
really bad surprises only. It is indeed a fact that exogenous shocks exist, as 
epitomized by the recent events of Sept. 11, 2001 and the coup in the 
Soviet Union on Aug. 19, 1991, which move stock market prices and create strong
bursts of volatility \cite{VolMRW}, as discussed above. However, it is
always the case? 
A key question is whether large losses and gains are indeed slaved to exogenous 
shocks or on the contrary may result from an endogenous origin in the dynamics 
of that particular stock market. The former possibility requires the risk 
manager to closely monitor the world of economics, business, political, social,
environmental news for possible instabilities. This approach 
is associated with standard ``fundamental'' analysis. The later endogenous 
scenario requires the investigation of signs of instabilities to be found in 
the market dynamics itself and could rationalize in part so-called 
``technical'' analysis (see \cite{andersen} and references therein).

\begin{figure}
\begin{center}
         \epsfig{file=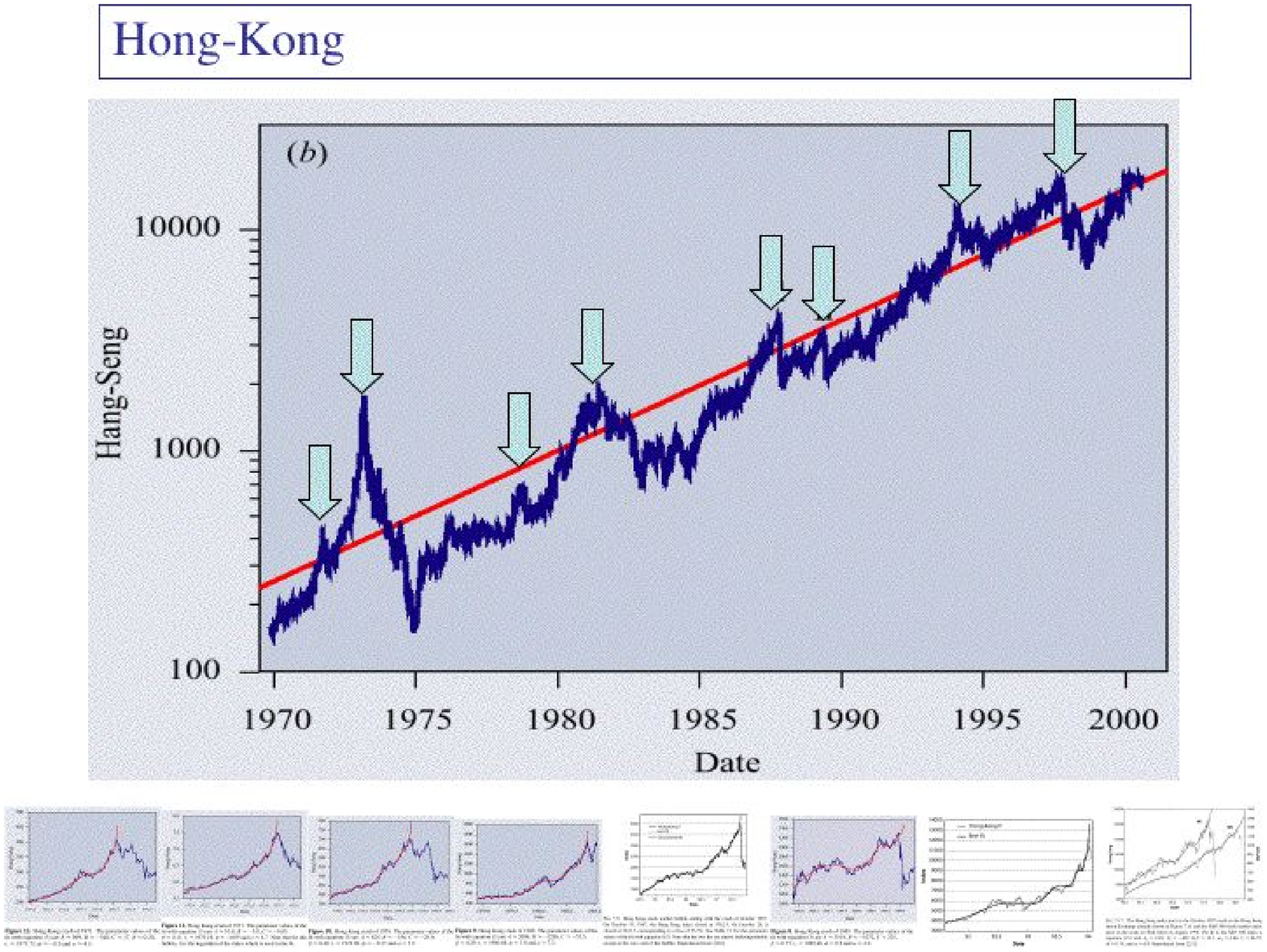,width=16cm,height=13cm}
     \caption{\label{crash_hk} The Hang-Seng composite index of the Hong Kong 
stock market from Nov. 1969 to Sept. 1999. Logarithmic scale in the vertical axis. 
The culmination
of the bubbles followed by strong corrections of crashes are indicated by the arrows
and correspond to the times Oct. 
1971, Feb, 1973, Sept. 1978, Oct. 1980, Oct. 1987, April 1989, Jan. 1994 and Oct. 1997. 
This figure shows that the Hang-Sing index has grown exponentially on average at the rate
of $\approx 13.6\%$ per year represented by the straight line corresponding to 
the best exponential fit to the data. Eight large bubbles (among them 5 are
very large) can be observed 
as upward accelerating deviations from the average exponential growth characterized
by LPPL signatures ending in a crash, here defined as a more than 15\% drop in less
than two weeks. The eight small panels at the bottom are given to show the LPPL
price trajectory over a period of six months preceding each of these 8 crashes.
Constructed from \protect\cite{SJ2001} and other papers from the author.}  
     \end{center}
\end{figure}

Johansen and Sornette \cite{JSendo} have carried out a systematic
investigation of crashes to clarify this question. They have proceded in
several steps. 
\begin{enumerate}
\item They have developed a methodology to identify crashes
as objectively and unambiguously as possible. Specifically, they have
studied the distribution of drawdowns (runs of losses) in several
markets:
the two leading exchange markets (US dollar against the Deutsch and against 
the Yen), the major world stock markets, the U.S. and Japanese
bond market and in the gold market. By introducing and varying a certain degree
of fuzziness in the definition of drawdowns, they have tested the robustness
of the empirical distributions of drawdowns.

\item By a careful analysis of these distributions, they have shown that 
the extreme tail belongs to a different population than the bulk (typically
the 1\% most extreme drawdowns occur 10 to 100 times more often than
would be predicted by an extropolation of the distribution of the 99\% 
remaining drawdowns). 

\item These extreme events which seem to belong to a different population
have been called ``outliers'' \cite{outl1,SJ2001,outl2,crashcom}. Others have refered to 
such events as ``kings'' or ``black swans.'' Johansen and Sornette \cite{JSendo}
have taken these kings as the crashes that need to be explained. Note that
this procedure ensures that the definition of a crash is relative to the 
specific market rather than obeying so arbitrary absolute rule.

\item Then, for each identified king, Johansen and Sornette \cite{JSendo} have checked whether
a specific market structure, called log-periodic power law (LPPL), 
is present in the price trajectory {\it preceding} the occurrence of the drawdown king.
The rational for this approach was based on their
previous works \cite{JS1999,JSL1999,JLS2000,SJ2001}, in which they 
documented that the existence of such
log-periodic power law signatures associated with speculative 
bubbles before crashes. The work \cite{JSendo} is in this respect
an out-of-sample test of the LPPL bubble-crash hypothesis applied
to a population of financial time series selected according to a
criterion (outlier test in the distribution of drawdowns)
which is unrelated to the LPPL structure itself. 

\item In this test, Johansen and Sornette \cite{JSendo} 
take the existence of a LPPL as the qualifying signature
for an endogenous crash: a drawdown outlier is seen as the 
end of a speculative unsustainable accelerating bubble generated endogenously. 

\item With these criteria fixed, Johansen and Sornette \cite{JSendo} identify
two classes of crashes. Those which are not preceded by a LPPL price trajectory
are classified as exogenous. It turns out that for those, it was possible to 
identify what seems to have been the 
relevant historical event, {\it i.e.}, a new piece of information of such 
magnitude and impact that it is reasonable to attribute the crash to it, 
following the standard view of the efficient market hypothesis. Such 
drawdown outliers are classified as having an exogenous origin. 

\item The second class, characterized by LPPL price trajectories, is called endogenous.
Figure \ref{crash_hk} illustrate a series of endogenous crashes preceded by LPPL
bubble trajectories on the Heng-Seng composite index of the Hong-Kong stock market, 
perhaps one of the most speculative markets in the world. All the events shown
belong to the endogenous class.

\item Globally over 
all the markets analyzed, Johansen and Sornette \cite{JSendo} 
identified 49 outliers, of which 25 were classified 
as endogenous, 22 as exogenous and 2 as associated with the Japanese 
``anti-bubble'' starting in Jan. 1990. Restricting to the world market indices,
they found 31 outliers, of which 19 are endogenous, 10 are exogenous and 2 are 
associated with the Japanese anti-bubble. 
\end{enumerate}

The combination of the two proposed 
detection techniques, one for drawdown outliers and the second for LPPL signatures, 
provides a novel and systematic taxonomy of crashes further substantiating the 
importance of LPPL (see also \cite{sorcrashbook,PRcritmar,zs1,sz1,ds2} for reviews and extensions).

A more microscopic approach formulated in terms of agent-based models has
also allowed to identify some mechanisms for the occurrence of 
extreme events, such as the excess bias on
nodes in the de Bruijn diagram of active agent strategies \cite{Johnbook}, or the decoupling 
of strategies which become transiently independent from the recent past \cite{vittig}.

\section{Concluding remarks}

Let us end by a discussion of other domains of applications.

While the idea is not yet developed, I think that beyond the products
sold by e-retailers discussed above, which are proxies of reputation and commercial
successes, the endo-exo question is relevant to understanding the
characteristics of Initial Public Offerings (IPO) \cite{ipo} and the
movie industry \cite{devany}. In the later, the mechanism of information
cascade derives from the fact that agents can observe box office
revenues and communicate word of mouth about the quality of the movies
they have seen.

Earthquakes are now thought to be due to a mixture of spontaneous occurrences driven
by plate tectonics and triggering by other previous earthquakes. Within such a picture
\cite{ETAS}
which rationalize many of the phenomenology of seismic catalogs, Helmstetter and Sornette
have shown that there is a fundamental limit to earthquake predictability resulting
from the ``exogenous'' class of earthquakes which are not triggered by other earthquakes
\cite{HSpredict}. Furthermore, the rate of foreshocks preceding mainshocks can 
be understood from the idea that mainshocks may result from endogenous
triggering by previous events, as developed above in section \ref{theo}. The 
dependence with time of the seismic rate of foreshocks
is predicted and observed to follow (\ref{mvmlw}). The memory kernels
$\phi(t)$ given by (\ref{mgmdfl}) and $K(t)$ given by (\ref{gmlasa}) correspond in the
present case respectively to the bare and renormalized Omori 
law \cite{Omori} for triggered aftershocks \cite{SS99,ETAS}.

The weather and the climate also involve extremely complex 
processes, which are often too difficult to disentangle. This leads
to major uncertainties in what are the important mechanisms that need
to be taken into account, for instance, to forecast the future global
warming of the earth due to anthropogenic activity coupled with natural
variability. 9/11 has again offered a unique window. Travis and Carleton \cite{Travis}
noted the following: ``Three days after
suicide airplane hijackers toppled the World Trade Center in New York
and slammed into the Pentagon in Washington, D.C., the station crew
noted an obvious absence of airborne jetliners from their perch 240
miles (384 kilometers) above Earth. I'll tell you one thing that's
really strange: Normally when we go over the U.S., the sky is like a
spider web of contrails, U.S. astronaut and outpost commander Frank
Culbertson told flight controllers at NASA's Mission Control Center in
Houston. `And now the sky is just about completely empty. There are no
contrails in the sky,' he added. `It's very, very weird.' `I hadn't
thought of that perspective,' fellow astronaut Cady Coleman replied.''
Travis and Carleton \cite{Travis} showed that a significant elevation
of the average diurnal temperature of the US in the three days following
9/11 when most jetliners were grounded and no contrails
were present. This is the archetype of 
an exogenous response. It remains to be seen if the endo-exo view point 
turns out to offer new fruitful perspectives to make progress in 
understanding and in forecasting the weather and the climate.

Finally, from a theoretical view point, another potentially 
interesting domain of research is to extend the 
concept of the response function to nonlinear systems \cite{NLPotter,mukaNL}
and to study its relationship with the internal fluctuations \cite{ruelle}.

{\bf Acknowledgments}: I am grateful to my collaborators and colleagues who
helped shape these ideas, among them, Y. Ageon, J. Andersen, R. Crane, D. Darcet,
F. Deschatres, T. Gilbert, S. Gluzman, A. Helmstetter, A. Johansen, Y.
Malevergne, J.-F. Muzy, V.F. Pisarenko, B. Roehner and W.-X. Zhou.

\end{document}